\documentclass[aps,pra,twocolumn,floatfix,superscriptaddress,nofootinbib,showpacs,a4paper,balancelastpage,twoside]{revtex4-2}
\usepackage{graphicx, amsmath,amssymb,psfrag,empheq,hyperref,natbib}
\usepackage{dcolumn}
\usepackage{bm}
\usepackage{epsfig}
\usepackage{xcolor}
\usepackage{amsfonts}
\usepackage{graphicx} 
\graphicspath{{}{../Bilder/}}
\usepackage{wasysym}
\usepackage{float}
\usepackage{siunitx}
\usepackage{hyperref}
\usepackage[paperwidth=210mm,paperheight=297mm,centering,hmargin=1.8cm,vmargin=2.5cm]{geometry}

\newcommand{\qbounce}{{\it{q}}{\sc{Bounce}}}				
\newcommand{\cannex}{{\sc{Cannex}}}				
\begin{document}
\title{Search for environment-dependent dilatons\\*\emph{Preprint Version}} 
\author{Hauke Fischer, Christian Käding, René I.P. Sedmik, Hartmut Abele}
\affiliation{Atominstitut, Technische Universität Wien, Stadionallee 2, A-1020 Vienna, Austria}
\author{Philippe Brax}
\affiliation{Institut de Physique Théorique, Université Paris-Scalay, CEA, CNRS, F-91191 Gif/Yvette Cedex, France}
\author{Mario Pitschmann}
\email{mario.pitschmann@tuwien.ac.at}
\affiliation{Atominstitut, Technische Universität Wien, Stadionallee 2, A-1020 Vienna, Austria}
%
\begin{abstract}
The environment-dependent dilaton field is a well-motivated candidate for dark energy and naturally arises in the strong coupling limit of string theory. In this article, we present the very first experimental constraints on the parameters of this model. For this, we employ data obtained from the \qbounce{} collaboration and the Lunar Laser Ranging  (LLR) experiment. Furthermore, we forecast expected exclusion plots for the Casimir And Non Newtonian force EXperiment (\cannex) soon to be realised in an improved setup. Finally, we provide a detailed analysis of the screening mechanism and additional symmetries of the dilaton field theory.\nopagebreak
\end{abstract}
\maketitle
%
\section{Introduction}

The origin of dark energy is one of the greatest puzzles in modern physics. Unexpectedly, type Ia supernovae data have revealed that our Universe is currently expanding at an accelerated rate~\cite{SupernovaCosmologyProject:1997zqe, SupernovaSearchTeam:1998fmf, SupernovaSearchTeam:1998bnz}. This has been confirmed by many other cosmological probes.

The theoretical framework describing the Universe on cosmological scales is general relativity (GR). As GR is a crucial ingredient in the interpretation of  cosmological observations,  it seems natural that modifying GR  could be at the heart of the  observed accelerated expansion of the Universe. While a modification at short distances is indeed easily realisable by extending the Einstein-Hilbert action with quantities invariant under general coordinate transformations and containing higher derivatives of the metric (see e.g.~\cite{Donoghue:1994dn}), a modification for large distance scales by making the theory massive is very intricate~\cite{deRham:2014zqa}. Amending GR by the so-called cosmological constant $\Lambda$ allows one to describe the accelerated expansion. However, such a procedure would lead to a severe fine-tuning problem~\cite{Sola:2013gha}. Consequently, the existence of new hypothetical scalar fields has been postulated, which couple to gravity and can account for dark energy~\cite{Joyce:2014kja}. Those new scalars generically lead to new interactions, so-called fifth forces and are theoretically well-motivated irrespective of their role for dark energy. As they have  avoided detection in past fifth force experiments, they must be subject to a  screening mechanism. Several such screening mechanisms have been devised, such as the chameleon~\cite{Khoury:2003rn,Khoury:2013tda}, K-mouflage~\cite{Brax:2012jr,Brax:2014wla}, Vainshtein~\cite{Vainshtein:1972sx} and Damour-Polyakov mechanisms~\cite{Damour:1994zq}. 

In this article, we investigate the dilaton model with a Damour-Polyakov mechanism. This is a screened scalar field model whose behaviour in local tests of gravity has been less studied so far~\cite{Damour:1994zq, Gasperini:2001pc,Damour:2002mi,Damour:2002nv,Brax:2011ja,Hartley:2018lvm,Kading:2023mdk}. This model has been proposed as a possible candidate for dark energy~\cite{Brax:2010gi,Sakstein:2014jrq}. Its potential naturally arises in the strong coupling limit of string theory and gives rise to a screening mechanism in connection with the Damour-Polyakov mechanism. Due to its origin in string theory this model is particularly well-motivated in comparison to similar models such as chameleons and symmetrons (for a related investigation concerning symmetrons we refer to~\cite{Cronenberg:2018qxf,Brax:2017hna,Pitschmann:2020ejb}). 

Herein, we provide a brief summary of this model, discuss its screening mechanism and parameter symmetries, followed by succinct descriptions of the corresponding experiments and methods that we employ in order to constrain the parameters of the dilaton. 
This article complements the theoretical analysis presented in~\cite{Brax:2022uyh}. 

\section{The dilaton with Damour-Polyakov mechanism}

The effective potential of the dilaton  is given by~\cite{Brax:2018iyo}
\begin{align}
V_{\text{eff}}(\phi; \rho) = V_0\,e^{- \lambda \phi / m_{\text{pl}}}+    \beta(\phi)\,\frac{\rho}{2m_{\text{pl}}}\,\phi\>, \label{Defdil}
\end{align}
where $V_0$ is a constant energy density, $\lambda$ a dimensionless constant, $\beta(\phi) = A_2 \phi /m_{\text{pl}}$ the full coupling to the matter density $\rho$, $A_2$ a dimensionless coupling constant and $m_{\text{pl}}$ the reduced Planck mass. Inside matter with density $\rho$, the dilaton field approaches its minimum value given by
\begin{align}
\phi_{\rho} = \frac{m_{\text{pl}}}{\lambda}\, W\left(\frac{\lambda^2 V_0}{A_2 \rho}\right)\>, \label{minimum}
\end{align}
where $W(x)$ is the Lambert $W$ function, which is the inverse function of $xe^x$.

This potential is motivated from the string dilaton $\chi$ and the condition $V(\chi) \rightarrow 0$ for $\chi \rightarrow \infty$, which is associated with the strong coupling limit of string theory~\cite{Gasperini:2001pc}. Hence, an asymptotic expansion $V(\chi) = \Tilde{V}_0\,e^{- \chi } + \Tilde{V}_1\,e^{- 2\chi }\ldots$ is applied. Furthermore, in Ref.~\cite{Damour:1994zq} it has been assumed that the coupling to matter has a minimum at some large value $\chi=\chi_0$. Consequently, near the minimum the coupling is proportional to $(\chi-\chi_0)^2$. Redefining $\phi:=\frac{m_{\text{pl}}}{\lambda}(\chi-\chi_0)$ leads to Eq.~(\ref{Defdil}) (for a full derivation see e.g.~\cite{MarioHabil}). 
For the derivation of experimental limits we demand the condition  
\begin{align}
A_2 \phi^2/(2 m_{\text{pl}}^2) \ll 1\label{cutoff}
\end{align}
to hold in order to ensure that couplings to matter of higher order in $\phi$ can be neglected.

The parameter space of this model can naturally be divided into three regions (see Appendix~\ref{AppSM1}).
A large enough $\lambda$ (at fixed $V_0$ and $A_2$) guarantees $e^{-\lambda \phi / m_{\text{pl}}} \ll 1$ and condition (\ref{cutoff}).
Inside this region the dilaton field primarily screens by increasing its mass in dense environments. Additionally, there is an approximate symmetry between $A_2$ and $V_0$; physical effects mainly depend on the product $A_2\, \text{ln}(V_0/\rho)$, but not on the individual values of $V_0$ and $A_2$ (see Appendix~\ref{AppSM2}). This is evident in the obtained experimental limits in Fig. (\ref{fig:OLD}) that shift systematically towards lower values of $A_2$ for increasing $V_0$. Condition (\ref{cutoff}) results in an ever stronger cut in the parameter space for increasing values of $V_0$ (for the calculation of limits a second cut-off was set to ensure that treating the experimental setups as 1D is appropriate).

\begin{figure*}[!ht]
\setcounter{figure}{1}
\centering
\includegraphics[width=\textwidth]
{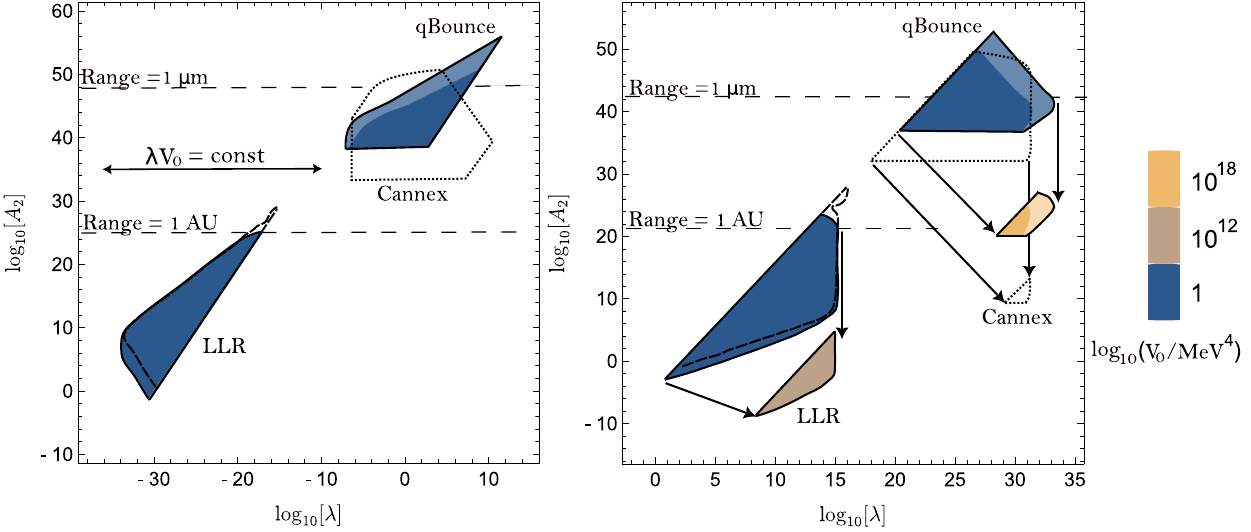}
\caption{The exclusion plots typically separate into two distinct regimes; \textit{Left}: here limits for small values of the parameter $\lambda$ are plotted; \textit{Right}: exclusion limits for large $\lambda$ are depicted (for further explanations we refer to the main text);\\*
\underline{LLR}: exclusion plots are filled areas in the bottom left in each region; limits obtained from violations of the equivalence principle are surrounded by solid lines, while limits from the precession of the lunar perigee are encompassed by dashed lines; \\*
\underline{\qbounce{}}: exclusion plots are  filled areas in the top right in each region; lighter areas correspond to micron screening and darker ones to fermi screening;  \\*
\underline{\cannex}: prospective limits are surrounded by dotted lines; the two areas plotted \textit{right} correspond to $\log_{10} (V_0 / {\rm MeV}^4) = 1$ and $10^{24}$, respectively.\label{fig:OLD}}
\end{figure*} 

If $\lambda$ is small enough then $e^{-\lambda \phi / m_{\text{pl}}}\simeq1$  and (\ref{cutoff}) holds. The dilaton field has a functional dependence only on the product of parameters $V_0 \lambda$ rather than on the individual parameters $V_0$ or $\lambda$, and the screening of the field in this region is primarily due to the decrease of its matter coupling $\beta(\phi)$ in dense environments (see Appendix~\ref{AppSM1}). Hence, computed limits in Fig. (\ref{fig:OLD}) simply shift towards lower values of $\lambda$ for increasing values of $V_0$ without changing their shapes, as long as $\lambda V_0$ is kept constant.

In between these two regions, for intermediate values of $\lambda$, there is a region where $A_2 \phi^2/(2 m_{\text{pl}}^2) \gg 1$ and, consequently, this effective dilaton model is outside its range of applicability. However, for $V_0 \ll 1$ MeV$^4$ the distinct experimental limits in Fig. (\ref{fig:OLD}) merge. The merging point depends on the specific experiment and on the vacuum density employed, but is qualitatively at $V_0 \sim 10^{-20}$ MeV$^4$. For much lower values of $V_0$ physical effects become weak and all experimental limits quickly disappear.

Tabletop experiments in a vacuum chamber play an important role in the search for screened scalar fields such as the dilaton. This follows from the low matter density within the vacuum chamber ensuring that the scalar field is less suppressed there than in dense environments, while sufficiently thick chamber walls effectively shield any influences from the outside world for a large region of parameter space. 
The same techniques have been utilised previously for experimental searches for chameleons~\cite{Jenke:2014yel, Cronenberg:2015bol} and symmetrons~\cite{Cronenberg:2018qxf}. Furthermore, screened scalar fields with comparably small interaction ranges can be probed better with tabletop experiments than via astrophysical searches.

\section{The \qbounce{} experiment}

In \qbounce{}~\cite{PhysRevD.81.065019,Jenke:2011zz,Jenke:2014yel} ultracold neutrons, which are totally reflected from most materials, are bouncing in the gravitational field of the Earth. The discrete energy levels are not equidistant, which allows to perform resonance spectroscopy in the gravitational field. 
In its realization corresponding to a Rabi setup~\cite{Cronenberg:2015bol}, neutrons pass through three regions: The first region acts effectively as a state selector and has a length of around 15 cm. A polished mirror at the bottom and a rough scatterer at a height of 20 $\mu$m on top ensure that only neutrons in the lowest few states can pass. Unwanted higher energy states are scattered out of the system. In the second region, neutrons pass a vibrating mirror with tunable frequency $\omega$ that can drive the neutron towards a higher energy state. This region has a length of 20 cm. The final region is identical to the first region (see Fig.~\ref{fig:qBounce} for a schematic setup). 
\begin{figure}[H]
\setcounter{figure}{0}
\begin{center}
\includegraphics[width=0.45\textwidth]
{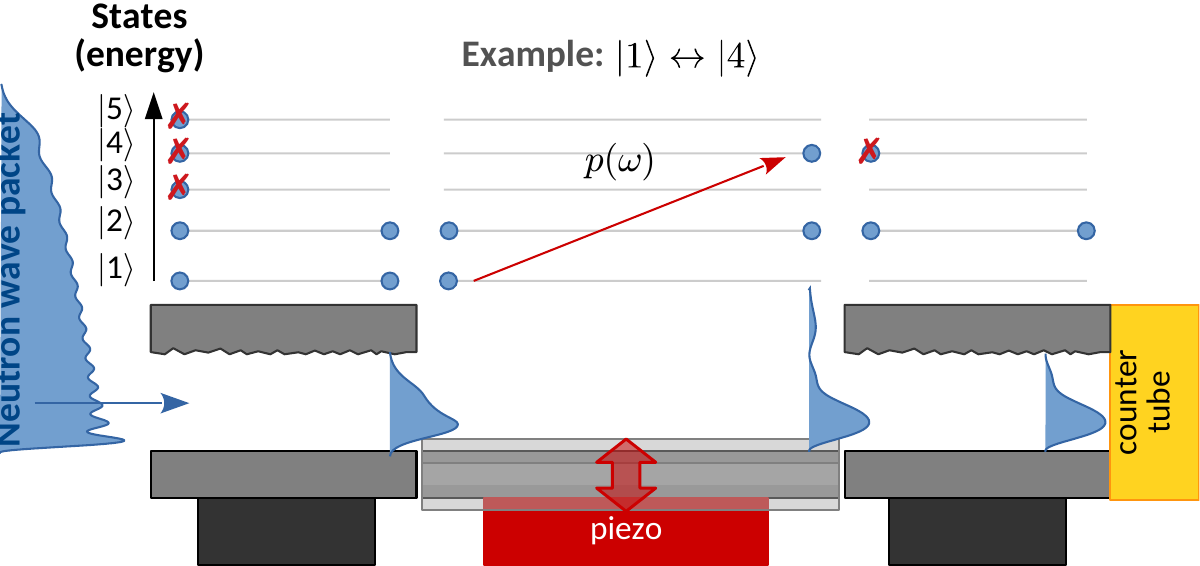}
\caption{Sketch of \qbounce{}}
\label{fig:qBounce}
\end{center}
\end{figure} 
If the energy $\hbar \omega$ associated with the frequency of the mirror is close to the energy $\Delta E_n = E_n - E_1$ needed to drive the neutron to a specific  higher energy state, the system enters a coherent superposition of the ground state and this excited state. If the neutron is not in one of the lowest $\sim 2$  states anymore when entering the last region, a loss in transmission is observed.

Since neutrons are electrically neutral and have very low polarizability, they are very insensitive to experimental background disturbances. Hence, \qbounce{} is a highly sensitive probe for new physics and has already been used to probe and set stringent limits on many hypothetical new interactions~\cite{Sponar:2020gfr}. Here, \qbounce{} is employed for the first time to set limits on the dilaton field. The presence of the latter would induce energy shifts that can directly be obtained from the stationary Schr\"odinger equation. Due to the comparatively large extension of the mirrors, the setup can safely be approximated as one dimensional, in which case the stationary Schr\"odinger equation reads
\begin{align}
&\bigg[-\frac{\hbar ^2}{2 m} \frac{\partial^2}{\partial z^2}+
  m g z  + \mathfrak{Q}\,\frac{A_2}{2}\frac{m}{m_{\text{pl}}^2}\,\phi^2(z)\bigg]\,\Psi_n(z) \nonumber\\  & = E_n\Psi_n(z)\>.
\end{align}
In general, this is a two-body problem since the mirror as well as the neutron interact with the dilaton field. We approximate this problem by treating the neutron as a sphere and extracting a ``screening charge" $\mathfrak{Q}$, which multiplies the dilaton potential and approximately describes the interaction of the neutron with the dilaton. For further details and an explicit expression for $\mathfrak{Q}$ we refer to the accompanying article~\cite{Brax:2022uyh}. Two limiting cases are considered depending on whether the neutron is described as a sphere of radius 0.5 fm in agreement with QCD (``fermi screening") or 5.9 $\mu$m corresponding to the natural extend of the wave function (``micron screening"). We assume that the true coupling lies within the boundaries provided by these two limiting cases.

For the calculation of the dilaton-induced energy shift, perturbation theory, as has been detailed in~\cite{Brax:2022uyh}, is not applicable for a large part of the parameter space since the computed effects of the dilaton field can be very large. Therefore, the eigenvalue problem associated with the stationary Schr\"odinger equation has been solved numerically to allow for a non-perturbative treatment. Details on this procedure can be found in Appendix~\ref{AppSM5}.

The experimental sensitivity achieved in the Rabi-like setup corresponds to an energy resolution of $\Delta E = 2 \times 10^{-15}$ eV in a vacuum chamber with a pressure of $2 \times 10^{-4}$ mbar. This sensitivity allows us to exclude a large part of the 3D parameter space of the dilaton field as shown in Fig.~\ref{fig:OLD}.

\section{Lunar Laser Ranging}

Lunar Laser Ranging (LLR) measures the distance between the surfaces of the Earth and the Moon with high precision. This method involves firing a laser beam at a retroreflector array installed on the lunar surface during the Apollo missions. The retroreflectors consist of a series of small mirrors that reflect the laser beam back to  Earth~\cite{muller2019lunar}.

Measuring the time it takes for the laser pulse to propagate to the Moon and back provides the distance between the two bodies with an accuracy of a few centimeters. This data has been used to measure the Moon's orbit to high experimental precision, which allows to test GR and set stringent limits on any alternative theories. To date, the data is compatible with GR, which necessitates that scalar fields with a non-minimal coupling to matter, if they exist, must have a screening mechanism.

Lunar Laser Ranging has been used to test the equivalence principle. Similarly, deviations from the inverse-square law of gravity would induce shifts in the precession of the lunar perigee. The experimental constraint for equivalence principle violations of the Earth (\earth) and Moon (\leftmoon) in the field of the  Sun (\sun) is given by~\cite{Brax:2022uyh,nordtvedt2001lunar}
\begin{align}
\delta_{\text{em}}\simeq \frac{|\vec{a}_{\phi \earth}-\vec{a}_{\phi \leftmoon}|}{|\vec{a}_G|}  \leq 2 \times 10^{-13}\>,
\end{align}
where $\vec{a}_\phi$ refers to the dilaton-induced acceleration towards the Sun in addition to the Newtonian acceleration $\vec{a}_G$.
A second constraint is placed on any shift of the precession of the lunar perigee given by
\begin{align}
\left|\frac{\delta \Omega}{\Omega}\right| &\simeq \left| \frac{R^2}{G M_{\earth}}\,\big(\delta f(R) + \frac{R}{2}\, \delta f'(R)\big)\right|\nonumber\\
& \leq 6.23833 \times 10^{-12}\>,
\end{align}
where $\delta f$ is the centripetal dilaton force per mass.
For the numerical generation of the corresponding dilaton limits we used the analytical results from Ref.~\cite{Brax:2022uyh}. The obtained exclusion volume is shown in Fig.~\ref{fig:OLD}.

\section{The \cannex{} experiment}

The Casimir And Non-Newtonian force EXperiment (\cannex{}) is currently being rebuilt at the Conrad Observatory in Austria~\cite{Sedmik:2021iaw}. It is especially designed to measure the Casimir force with unprecedented accuracy as well as fifth forces due to hypothetical new interactions, and gravity. The experimental setup consists of two plane parallel plates in close proximity, and allows to measure induced forces and their gradients between these plates in direct or Cavendish configuration (see Fig.~\ref{fig:NEW3} for a schematic setup). 
\begin{figure}[H]
\setcounter{figure}{2}
\begin{center}
\includegraphics[width=0.5\textwidth]
{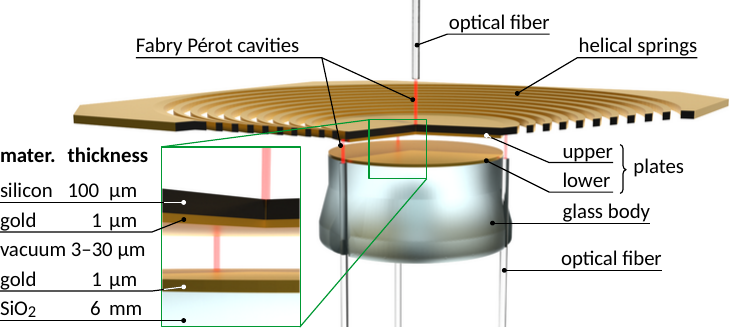}
\caption{Schematic cut view of the \cannex{} setup in direct configuration (without electrostatic shield between the plates). Forces are detected using Fabry P\'erot interferometers sensing the extension of the mass-spring system created by the helical springs and the upper plate. The insert on the left defines the material and thickness of the various layers. Note that the upper plate and the springs are coated in addition on all sides with a thin (\SI{50}{\nano\metre}) layer of gold.}
\label{fig:NEW3}
\end{center}
\end{figure}

Due to the geometry of two truly parallel plates, force generation by any interaction is maximized. With an effective area of \SI{1}{\centi\metre^2} and a targeted sensitivity of \SI{0.1}{\nano\newton/\metre^2} at separations between 3 and \SI{30}{\micro\metre}, the Casimir effect as well as several hypothetical interactions could be measured at unprecedented accuracy~\cite{Sedmik:2021iaw}.  
By varying the pressure of Xe gas, the vacuum density surrounding the plates can be tuned between 5.3$\times 10^{-12}$ kg/$\text{m}^3$ and 0.0026 kg/$\text{m}^3$.
This variability allows for relative measurements triggering the distinctive feature of hypothetical new scalar fields with non-minimal coupling to matter -- their strong sensitivity to ambient densities. \cannex{} therefore will be a powerful tool in the search for such interactions. 
In one dimension, the setup can approximately be modeled as a half space with density $\rho_M=\SI{2514}{\kilogram/\metre^3}$ for $z\leq -d$, a vacuum region with density $\rho_V$ for $-d<z<d$, an upper plate with density $\rho_M$ for $d<z<d+D$, and a vacuum region with density $\rho_V$ for $z>d+D$. The upper plate has a thickness of $D= \SI{100}{\micro\metre}$ and is movable, such that $\SI{1.5}{\micro\metre}<d<\SI{15}{\micro\metre}$. If dilatons indeed exist, they would induce an additional pressure between the plates. To compute this pressure, the corresponding differential equation for the dilaton field 
\begin{align}
\frac{d^2 \phi}{dz^2} + \frac{\lambda V_0}{m_{\text{pl}}}\,e^{-\lambda \phi/m_{\text{pl}}}-\frac{A_2\rho(z)}{m_{\text{pl}}^2}\,\phi =0\>,
\end{align}
has been solved numerically for all parameters of interest.

\begin{figure}[H]
\begin{center}
\includegraphics[width=0.4\textwidth]
{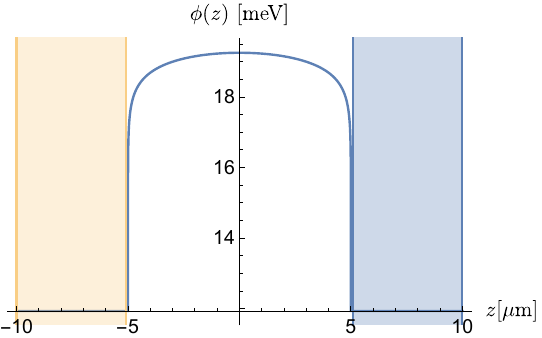}
\caption{Simulated dilaton field in between the parallel plates of the \cannex \ setup for $\lambda = 10^{31}, A_2 = 10^{45}$ and  $V_0 = 10$ MeV$^4$; the lower (yellow) and upper (blue) plates are indicated in color.}
\label{fig:NEW2}
\end{center}
\end{figure}

For further details on the simulations and the pressure calculation we refer to Appendices~\ref{AppSM3} and \ref{AppSM4}. An example of a simulated dilaton field for the \cannex{} setup is provided in Fig. \ref{fig:NEW2}.

\section{Dilaton dark energy}

\begin{figure}[H]
\begin{center}
\includegraphics[width=0.45\textwidth]
{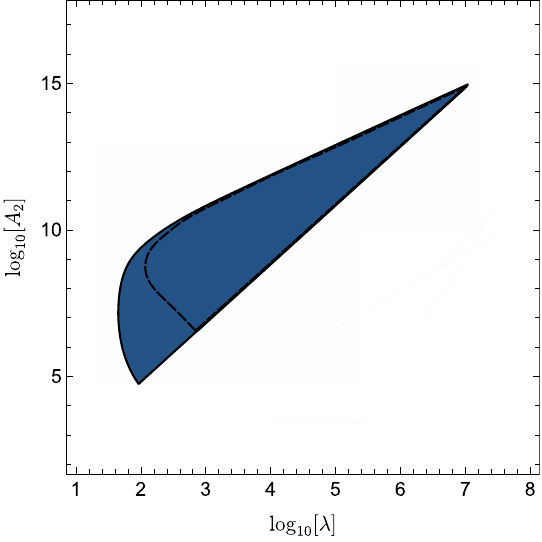}
\caption{Limits for the dilaton field as the source of dark energy. Only LLR can set limits in this case. Inside the plotted region $V_0$ takes the value $V_0 \simeq 3 \Omega_{\Lambda_0} m_{\text{pl}}^2 H_0^2$.}
\label{fig:Cosmos}
\end{center}
\end{figure}

Requiring that the dilaton provides the vacuum energy  accounting for dark energy results in a reduction of parameter space to two dimensions, where the condition $V_{\text{eff}}(\phi_V; \rho_V) = 3 \Omega_{\Lambda_0} m_{\text{pl}}^2 H_0^2 $ holds for the cosmological vacuum density $\rho_V$ with the corresponding field minimum $\phi_V$. This idea has been detailed in Ref.~\cite{Brax:2022uyh}, where it has been shown that $V_0$ can then be expressed in analytically closed form as a function of $\lambda$ and $A_2$. The numerical analysis shows that such dark energy dilatons violate condition (\ref{cutoff}) inside the entire parameter region where $e^{-\lambda \phi _V/ m_{\text{pl}}}\ll 1$ for the cosmological vacuum density $\rho_V$. Interestingly, $A_2 \phi_V^2/(2 m_{\text{pl}}^2) \sim  1$ is roughly constant in this region. The larger part of the experimentally feasible parameter space where $e^{-\lambda \phi_V/ m_{\text{pl}}}\simeq 1$ also violates condition (\ref{cutoff}). This is the reason why there are only comparably small excluded areas for lunar laser ranging (see Fig.~\ref{fig:Cosmos}), while there are no other limits for the tabletop experiments considered herein. 
However, if the dilaton field were to  contribute only 10\% or less to the dark energy,  condition (\ref{cutoff}) would pose no strong restrictions any more, which would allow to exclude large areas of the 2D parameter space for all investigated experiments in this case. 

\section{Discussion}

The analysis provided herein shows that LLR is sensitive to the dilaton field for interaction ranges in vacuum of approximately 1 AU and larger, while the tabletop experiments considered herein can probe the field for ranges as low as 1 $\mu$m in agreement with expectations. In the future, \cannex{} will be able to access a large part of the parameter space which is left open by \qbounce{} and LLR. If the dilaton is the only source of dark energy, only minor measurable effects are expected.

The code that has been used to generate all obtained results is available at \cite{H.Fischer_DilatonCode}.

\section{Acknowledgments}

This article was supported by the Austrian Science Fund (FWF): P 34240-N, P-33279-N, P36577-N, and is based upon work from COST Action COSMIC WISPers CA21106,
supported by COST (European Cooperation in Science and Technology).
We thank Tobias Jenke, measurements with \qbounce{} were performed at the ultra-cold beam position PF2@Institut Laue-Langevin, Grenoble.

\appendix
\section{Supplementary materials}

\subsection{Derivation of the three parameter regions, the screening mechanisms and the parameter symmetry}\label{AppSM1}

In this section, we describe the three regions of the parameter space obtained by varying the magnitude of $\lambda$.
Increasing $\lambda$ while keeping the other parameters fixed eventually leads to
\begin{align}
    \frac{\lambda^2 V_0}{A_2 \rho} \gg 1\>. \label{largarg}
\end{align}
Using $W(x)\simeq \text{ln}(x)- \text{ln} (\text{ln}(x))$ for large $x$ we can approximate 
\begin{align}
    \phi_{\rho} \simeq \frac{m_{\text{pl}}}{\lambda}\left\{\ln\left(\frac{\lambda^2 V_0}{A_2 \rho}\right)-\ln\left[\ln\left(\frac{\lambda^2 V_0}{A_2 \rho}\right)\right]\right\}, \label{fracfrac}
\end{align}
which shows that
\begin{align}
    e^{-\lambda \phi_{\rho} /m_{\text{pl}} } \simeq \text{ln}\left(\frac{\lambda^2 V_0}{A_2 \rho}\right)\bigg/\left(\frac{\lambda^2 V_0}{A_2 \rho}\right) \ll 1\>.
\end{align}
The mass $\mu_{\rho}$ of the dilaton is given by~\cite{Brax:2022uyh}
\begin{align}
\mu_{\rho}&=\frac{1}{m_{\text{pl}}}\,\sqrt{\lambda^2 V_0\,e^{- \lambda \phi_{\rho}/m_{\text{pl}}}+A_2\rho} \nonumber\\
&\simeq \frac{\sqrt{A_2 \rho}}{m_{\text{pl}}}\sqrt{1+ \ln\left(\frac{\lambda^2 V_0}{A_2 \rho}\right)} \nonumber\\
&\simeq \frac{1}{m_{\text{pl}}}\sqrt{A_2 \rho\,\ln\left(\frac{\lambda^2 V_0}{A_2 \rho}\right)}\>.  \label{mass}
\end{align}
Then, the full coupling to matter is approximately
\begin{align}
    \beta(\phi_{\rho}) &= \frac{A_2 \phi_{\rho}}{m_{\text{pl}}} \nonumber\\
    &\simeq \frac{A_2}{\lambda}\left\{\ln\left(\frac{\lambda^2 V_0}{A_2 \rho}\right)-\ln\left[\ln\left(\frac{\lambda^2 V_0}{A_2 \rho}\right)\right]\right\}.\label{beta}
\end{align}
Since $\rho$ effects $\beta(\phi_{\rho})$ only logarithmically (as long as Eq.~(\ref{largarg}) holds), while the mass has a square root dependence, increasing the density primarily leads to an increase in the mass of the field but only a negligible decrease of $\beta(\phi_\rho).$ 

Decreasing $\lambda$ inside this region increases $\phi_\rho$ according to Eq.~(\ref{fracfrac}), which eventually leads to a violation of the condition  $A_2 \phi^2/(2 m_{\text{pl}}^2) \ll 1$. Eventually, however, $\lambda$ gets small enough such that $\lambda^2 V_0/(A_2 \rho) \ll 1$ holds. Hence, using $W(x) \simeq x$ for small $x$, we obtain in this second region
\begin{align}
    \phi_{\rho} &\simeq m_{\text{pl}}\,\frac{\lambda V_0}{A_2 \rho}\>, \\
    e^{-\lambda \phi_\rho / m_{\text{pl}}} &\simeq e^{-\frac{\lambda^2 V_0}{A_2 \rho}} \simeq 1\>, \\
    \mu_{\rho} &\simeq \frac{\sqrt{A_2 \rho}}{m_{\text{pl}}}\>,\\
    \beta(\phi_{\rho}) &\simeq \frac{\lambda V_0}{\rho}\>.
\end{align}
Decreasing $\lambda$ inside this second region decreases $\phi_\rho$ (in contrast to the behaviour in the first region) and hence the condition $A_2 \phi^2/(2 m_{\text{pl}}^2) \ll 1$ is eventually fulfilled again. Inside this parameter region, $\beta(\phi_{\rho})$ decreases considerably by increasing $\rho$. Finally, since 
\begin{align}
    V_{\text{eff}}(\phi) &= V_0\, e^{- \lambda \phi /m_{\text{pl}}} + \frac{A_2\rho}{2 m_{\text{pl}}^2}\, \phi^2 \nonumber\\
    &\simeq V_0 - \lambda V_0\, \frac{\phi}{m_{\text{pl}}} + \frac{A_2\rho}{2 m_{\text{pl}}^2}\, \phi^2\>,
\end{align}
only the product of $\lambda V_0 $ enters the equations of motion, which explains the parameter symmetry that was observed also numerically, i.e. changing the parameters $\lambda$ and  $V_0$ whilst keeping their product $\lambda V_0$ fixed preserves the constraints on the parameter space for small enough $\lambda$.

\subsection{Additional explanation for the exclusion plots in the $e^{-\lambda \phi / m_{\text{pl}}} \ll 1 $ region}\label{AppSM2}

There is another approximate symmetry inside the $e^{-\lambda \phi / m_{\text{pl}}} \ll 1$ region, which explains why the exclusion plots shift systematically towards lower values of $A_2$ when increasing $V_0$. To leading order the parameters $A_2$ and $V_0$ enter the full coupling to matter (\ref{beta}) and the dilaton mass (\ref{mass}) via the same functional dependence $ A_2 \ln\left(\lambda^2 V_0 / A_2 \rho\right)$. In the excluded regions in the main paper $\ln\left(V_0 /\rho\right) \gg \ln\left(\lambda^2 /A_2\right) $ holds for essentially all of the displayed parameter space. Hence, $ A_2 \ln\left(\lambda^2 V_0 / A_2 \rho\right) \simeq A_2  \ln\left(V_0 / \rho\right) $. Therefore, the full coupling as well as the dilaton mass essentially depend only the product  $A_2  \ln\left(V_0 / \rho\right)$, which is why there is an approximate symmetry between these two parameters. Hence, increasing $V_0$ can effectively be compensated by a corresponding decrease of $A_2$ as has been observed in the excluded regions. In contrast, the precession of the lunar perigee does not follow that symmetry. This is due to the sum of two physical effects with opposite signs that cancel each other for larger values of $V_0$ in this case.

\subsection{Derivation of the pressure in the \cannex{} experiment}\label{AppSM3} 

For numerical calculations we made use of the formula for the pressure $P_z$ on the upper plate of the \cannex{} setup
\begin{align}
P_z = \frac{\rho_M}{\rho_M-\rho_V}\,\big(V_{\text{eff}}(\phi_V,\rho_V)-V_{\text{eff}}(\phi_0,\rho_V)\big)\>, \label{newpress}
\end{align}
where $\phi_0 = \phi(0)$ is the dilaton field value at the center between both plates 
and the effective potential is given by
\begin{align}
V_{\text{eff}}(\phi;\rho) = V(\phi) + \rho A(\phi)\>.
\end{align}
In \cite{Brax:2022uyh} the relation 
\begin{align}
P_z&=\rho_M\,\big(\ln A(\phi(d))-\ln A(\phi(d+D)\big) \nonumber\\
&\simeq \rho_M\,\big(A(\phi(d))-A(\phi(d+D)\big)\>, \label{press}
\end{align}
has been obtained, where in the second line $A(\phi)\simeq 1$ has been used, which holds for all models of interest as e.g. dilatons, symmetrons and chameleons.
However, this relation has been found challenging to work with numerically due to extreme slopes of the dilaton field near the mirror surfaces.  

Therefore, it turns out that the relation for the pressure  Eq.~(\ref{newpress})  is more convenient for  numerical simulations. We detail its derivation in what follows. Due to the screening mechanism, the field assumedly takes on its minimum value $\phi_{M}$ inside the upper mirror of thickness $D$  (this has been checked explicitly for all parameter values where limits have been set) and the value of $\phi(d)$ is therefore to a very good approximation given by the value at the surface of a two-mirror setup, where both mirrors are infinitely extended with a vacuum region in between them. Analogously, the value $\phi(d+D)$ is given by the value at the surface of the setup where one mirror is infinitely extended with a vacuum region above. 
In \cite{Brax:2022uyh} the integrated equation of motion  
\begin{align}
\frac{1}{2}\left(\frac{d \phi}{dz}\right)^2 - \frac{1}{2}\left(\frac{d \phi}{dz}\right)^2\bigg|_{z=z_0}  =  V_{\text{eff}}(\phi; \rho) -  V_{\text{eff}}(\phi; \rho)\big|_{z=z_0}\>, \label{ieom1}
\end{align}
has been derived. For the one-mirror case we take the boundary conditions $\phi(z)\rightarrow \phi_{M}$ for $z \rightarrow -\infty$ and $\phi(z)\rightarrow \phi_V$ for $z \rightarrow \infty$. In the limit $z \rightarrow \infty$ we get
\begin{align}
 - \frac{1}{2}\left(\frac{d \phi}{dz}\right)^2\bigg|_{z=z_0}  =  V_{\text{eff}}(\phi_V; \rho_V) -  V_{\text{eff}}(\phi; \rho)\big|_{z=z_0}\>. \label{ieom2}
\end{align}
Subtracting Eq.~(\ref{ieom2}) from Eq.~(\ref{ieom1}) gives inside the vacuum
\begin{align}
\frac{1}{2}\left(\frac{d \phi}{dz}\right)^2  =  V_{\text{eff}}(\phi; \rho_V) -  V_{\text{eff}}(\phi_V; \rho_V)\>.
\end{align}
Similarly, inside the mirror we find 
\begin{align}
\frac{1}{2}\left(\frac{d \phi}{dz}\right)^2  =  V_{\text{eff}}(\phi; \rho_M) -  V_{\text{eff}}(\phi_M; \rho_M)\>.
\end{align}
By continuity of the derivative at $z=d+D$ we straightforwardly obtain
\begin{align}
 A\big(\phi(d+D)\big) = \frac{1}{\rho_M-\rho_V}\,\big( V_{\text{eff}}(\phi_M; \rho_M)-V_{\text{eff}}(\phi_V; \rho_V)\big)\>. 
\end{align}
In case of the two infinitely extended mirrors we can use analogous reasoning, using that $\partial \phi / \partial z|_{z=0} =0$ due to the symmetry of the setup with $\phi_0:= \phi(0)$ being the value at the center between both mirrors. This results in
\begin{align}
 A(\phi(d)) = \frac{1}{\rho_M-\rho_V}\,\big( V_{\text{eff}}(\phi_M; \rho_M)-V_{\text{eff}}(\phi_0; \rho_V)\big)\>. \label{A1}
\end{align}
Substituting these results into Eq.~(\ref{press}) proves Eq.~(\ref{newpress}).

\subsection{Details of numerical simulations of the dilaton field for \cannex{}}\label{AppSM4}

For our numerical simulations we used Mathematica 13.1. We found that the built-in NDSolve function for solving differential equations numerically does not work well for simulating the dilaton field, or solving the Schr\"odinger equation in the presence of a dilaton field. Therefore, we wrote our own code adapted to solving these equations. We work with a non-uniform finite difference method to approximate the second derivative of $\phi$ occurring in both differential equations, namely \cite{tan1990self}

\begin{align}
\phi_{i}'' \approx \frac{2(\phi_{i+1}-\phi_{i})}{h_i(h_i+h_{i-1})}-\frac{2(\phi_{i}-\phi_{i-1})}{h_{i-1}(h_i+h_{i-1})}
\end{align}
 with $h_i := x_{i+1}-x_i$ and the non-uniform approximation of the simulation interval $x_1, ..., x_N$. For the one dimensional dilaton field this results in the discretized differential equation

\begin{align}
\frac{2(\phi_{i+1}-\phi_{i})}{h_i(h_i+h_{i-1})}-\frac{2(\phi_{i}-\phi_{i-1})}{h_{i-1}(h_i+h_{i-1})} + \nonumber\\
\frac{\lambda V_0}{m_{\text{pl}}}\,e^{- \lambda \phi_i/m_{\text{pl}}} - \frac{A_2}{m_{\text{pl}}^2}\,\rho_i \phi_i = 0\>.\label{discretizte}
\end{align}
This is a non-linear system of equations on $\mathbb{R}^N$ that we solved with a self-programmed Newton's method. Boundary conditions were implemented by setting $\phi_0 = \phi_{N+1} = \phi_M$. This allowed us to use an arbitrary mesh, which we fine-tuned for the  dilaton field profiles. Unlike the built-in finite element method that also allows arbitrary meshes, our algorithm is not restricted to machine precision, but works with arbitrary precision, which is a major advantage for the dilaton field. Furthermore, we found that the non-linear FEM algorithms in Mathematica often fail to converge to the correct solution without returning any error messages and are therefore unreliable. Our code is freely available for investigation of any further details at \cite{H.Fischer_DilatonCode}.

\subsection{Details for computing the energy shifts for \qbounce{}}\label{AppSM5}

We used perturbation theory when applicable. In all other cases we discretized the Hamilton operator using the same discretization method as explained for \cannex{}. The corresponding discretized version of the stationary Schr\"odinger equation is hence given by

\begin{align}
-\frac{1}{2m}\left[\frac{2(\Psi_{i+1}-\Psi_{i})}{h_i(h_i+h_{i-1})}-\frac{2(\Psi_{i}-\Psi_{i-1})}{h_{i-1}(h_i+h_{i-1})}\right] + V_i\Psi_{i} = E \Psi_{i} \label{FDS}
\end{align}
with $\displaystyle V_i = \mathfrak{Q}\,\frac{A_2}{2}\frac{m_N}{m_{\text{pl}}^2}\,\phi^2(x_i)$.
 This results in a discrete approximation of the Hamilton operator given by

\begin{align}
H_{ij} = \begin{cases}
      -\displaystyle\frac{1}{2m} \frac{2}{h_i(h_i+h_{i-1})} &\text{, if } j = i+1\\
     \displaystyle-\frac{1}{2m} \frac{2}{h_{i-1}(h_i+h_{i-1})} &\text{, if } j = i-1\\
     -H_{ii} - H_{i,i-1} + V_i&\text{, if } j = i \\
     0 &\text{, else}\>.
    \end{cases} 
\end{align}
 Boundary conditions can be implemented analogously to the dilaton field simulation. Since the resulting approximation for the Hamilton operator is not symmetric on non-uniform grids, in our code we applied a transformation to restore symmetry following \cite{tan1990self}. This procedure results in an eigenvalue problem for a $N \times N$ matrix that can easily be solved numerically, and returns all possible eigenstates and eigenvalues obtainable with the fineness of the grid, from which we can safely extract the first and fourth energy state, and the corresponding energies. Due to the high computational cost of this procedure, we only computed around 10 points for the remaining non-trivial edge (which does not come from a cut-off or can be obtained from perturbation theory) of the exclusion area and fitted the result with a linear function, which approximates the contour well. This procedure is justified because the difference between fermi and micron screening, which is our error guess of the edge of the exclusion area, is much larger than the error introduced by our fit.

\bibliography{refs}
\end{document}